\documentclass[prl,aps,superscriptaddress,floatfix,nofootinbib,twocolumn]{revtex4}
\usepackage{exscale}                  
\usepackage[intlimits]{amsmath}       
\usepackage{amsfonts}
\usepackage{amssymb,amscd}
\usepackage[dvips]{epsfig}                   
\usepackage{array}
\usepackage{wrapfig}
\usepackage{color}
\newcommand{\is}{\sum\!\!\!\!\!\!\!\int}	
\newcommand{\isline}{{\scriptstyle \sum}\!\!\!\!\!\int\:\!}	
\newcommand{\beqa}{\begin{eqnarray}}
\newcommand{\eeqa}{\end{eqnarray}}
\newcommand{\beq}{\begin{equation}}
\newcommand{\eeq}{\end{equation}}

\begin{document}
\title{Chiral and deconfinement phase transitions of two-flavour QCD\\
 at finite temperature and chemical potential}
\author{Christian~S.~Fischer}
\affiliation{Institut f\"ur Theoretische Physik,
  Justus-Liebig-Universit\"at Gie\ss{}en,
  Heinrich-Buff-Ring 16,
  D-35392 Gie\ss{}en, Germany}
\affiliation{GSI Helmholtzzentrum f\"ur Schwerionenforschung GmbH, 
  Planckstr. 1  D-64291 Darmstadt, Germany.}
\author{Jan Luecker}
\affiliation{Institut f\"ur Theoretische Physik,
  Justus-Liebig-Universit\"at Gie\ss{}en,
  Heinrich-Buff-Ring 16,
  D-35392 Gie\ss{}en, Germany}
\author{Jens~A.~Mueller}
\affiliation{Institut f\"ur Kernphysik, 
  Technische Universit\"at Darmstadt,
  Schlossgartenstra{\ss}e 9,\\ 
  D-64289 Darmstadt, Germany}
\date{\today}

\begin{abstract}

We present results for the chiral and deconfinement transition of two flavor 
QCD at finite temperature and chemical potential. To this end we study the
quark condensate and its dual, the dressed Polyakov loop, with functional
methods using a set of Dyson-Schwinger equations. 
The quark-propagator is determined self-consistently within a truncation scheme
including temperature and in-medium effects of the gluon propagator.
For the chiral transition we find a crossover 
turning into a first order transition at a critical endpoint at large quark chemical 
potential, $\mu_{EP}/T_{EP} \approx 3$. For the deconfinement 
transition we find a pseudo-critical temperature above the chiral transition 
in the crossover region but coinciding transition temperatures close to the 
critical endpoint.

\end{abstract}

\maketitle

{\bf Introduction}\\
Heavy ion collision experiments at RHIC, the LHC 
and the future FAIR project are designed to study the properties of the large 
temperature phase of QCD, the quark-gluon plasma. 
Of particular interest are the details of the chiral and deconfinement transition.
Here, a standard scenario favors a chiral crossover at 
small chemical potential turning into a first order chiral transition at a critical 
point. The search for this critical point is one of the main motivations 
in the physics program of the future CBM/FAIR experiment in Darmstadt and the 
NICA project in Dubna. However, on the theoretical side there is ambiguity not 
only about the location of the critical point but also if the standard scenario
is even correct. Among other possibilities \cite{deForcrand:2008zi}
an exotic, quarkyonic matter phase \cite{McLerran:2007qj} or
inhomogeneous chiral condensates \cite{Nickel:2009ke,Kojo:2009ha} could
be realized. Unfortunately, lattice Monte-Carlo simulations cannot be used 
to resolve this issue due to the notorious sign problem.

Effective models to QCD like the Polyakov-loop extended Nambu--Jona-Lasinio
model (PNJL) \cite{Fukushima:2003fw} and the Polyakov-loop extended 
quark-meson model (PQM) \cite{Schaefer:2007pw,Herbst:2010rf}
constitute an alternative approach. Much progress has been made taking into 
account fluctuations beyond mean-field \cite{Herbst:2010rf}, which  
shift for example the critical endpoint to larger chemical potential.
Interestingly this is in agreement with recent lattice results indicating that the 
critical endpoint if existent is at much larger chemical potential \cite{Endrodi:2011gv}
than previously thought \cite{lattice}.

Within QCD fortunately also non-perturbative tools not limited by the
sign problem are at our disposal. In particular, it has been shown that 
Dyson-Schwinger equations (DSEs) and the functional 
renormalization group \cite{Roberts:2000aa,Braun:2006jd,Fischer:2009wc,Fischer:2010fx,
Braun:2009gm,Mueller:2010ah,Qin:2010nq} 
are capable to describe both, the chiral and the deconfinement 
transition at finite temperatures and imaginary chemical potential 
\cite{Fischer:2009wc,Braun:2009gm}.

In this letter we extend these applications to finite, real chemical potential. 
We use the DSEs for the propagators of Landau gauge QCD with two fermion flavors 
in a suitable truncation scheme. Working with physical quark masses we determine
the quark condensate and its dual, the dressed Polyakov loop in the
real $(T,\mu)$-plane. We extract 
transition temperatures for the chiral and deconfinement transition and determine 
their nature. We locate the critical endpoint of the chiral transition. Our study 
provides the tools for further, more detailed investigations of gauge invariant 
properties of the QCD phase diagram by functional methods.

\vspace*{1mm}
{\bf Order parameters}\\
We calculate the order parameters for chiral symmetry breaking and 
confinement from the fully dressed quark propagator, given by
\beq
S^{-1}(p) = i\vec{\gamma}\vec{p}A(p)+i\gamma_{4}(\omega_n+i\mu)C(p) + B(p),
\eeq
where $\mu$ is the quark chemical potential and $\omega_n = \pi T (2n+1)$ 
are the Matsubara modes in the imaginary time formalism with temperature $T$.
The argument $(p)$ serves as an abbreviation for $(\vec{p}^2,\omega_n)$.
The scalar and vector dressing functions $B$ and $A,C$ are calculated from 
the quark DSE. For the bare, renormalized propagator $S^{-1}_0(p)$ we have 
$A=C=Z_2$ and $B=m$, where $m$ is the renormalized bare quark mass
and $Z_2$ is the quark renormalization factor.

There are several equivalent choices of order parameters for chiral symmetry 
breaking; here we use the quark condensate
\beq \label{eq:condensate}
\langle\bar{\psi}\psi\rangle = 
Z_2T\sum_n\int\frac{d^3p}{(2\pi)^3}\mathrm{Tr}_D\left[S(p)\right].
\eeq
%
We work with two degenerate quark flavors with physical quark masses
and therefore expect a smooth crossover at vanishing chemical potential 
\cite{lattice}. Below, we determine the transition 
temperatures via  the maximum of the susceptibility 
$\frac{\partial \langle\bar{\psi}\psi\rangle}{\partial m}$.

Extracting an order parameter for confinement from the quark propagator is 
a much more challenging task. In the last years a method has been developed 
to calculate dual condensates $\Sigma_n$ which are sensitive 
to centre symmetry breaking \cite{Gattringer:2006ci,Synatschke:2007bz,Bilgici:2008qy}. 
They are defined by the transform
\beq
\Sigma_n = \int_\varphi e^{-i\varphi n} \langle\bar{\psi}\psi\rangle_\varphi,
\label{eq:dual}
\eeq
where $\int_\varphi \equiv \int_0^{2\pi}\frac{d\varphi}{2\pi}$ and $\langle\bar{\psi}\psi\rangle_\varphi$ 
is the quark condensate evaluated at generalized, $U(1)$-valued boundary conditions 
$\psi(\vec{x},1/T)=e^{i\varphi}\psi(\vec{x},0)$ with $\varphi \in [0,2\pi[$. 
On a lattice, the quantity $\Sigma_n$ can be interpreted 
as a sum of all closed loops winding $n$ times around the Euclidean time direction. 
Consequently, $\Sigma_{+ 1}$ contains the Polyakov loop and 
$\Sigma_{-1}$ its conjugate, together with all 
other oriented loops that wind once around
the space-time torus. They therefore have been 
called 'dressed Polyakov loop' and its conjugate \cite{Bilgici:2008qy}. For 
$\Sigma_{\pm 1}$ to act as order parameters for deconfinement it is mandatory to 
implement the generalized, $U(1)$-valued boundary conditions only on the level of 
observables, but not in the partition function itself. All 
closed quark loops therefore maintain the physical value $\varphi = \pi$, whereas 
$\varphi \in [0,2\pi[$ for the test quark in 
Eq.~(\ref{eq:dual}). This procedure breaks the Roberge-Weiss symmetry, a necessary 
condition for the dual condensate to act as an order parameter 
for centre symmetry breaking \cite{Braun:2009gm,Fukushima:2003fm}.

For our choice of physical quark mass we expect a
deconfinement crossover at small $\mu$. To determine
the pseudo-critical temperature we use the maximum of the 
derivative with respect to the quark mass.

\begin{figure}[t]
\centerline{\includegraphics[width=\columnwidth]{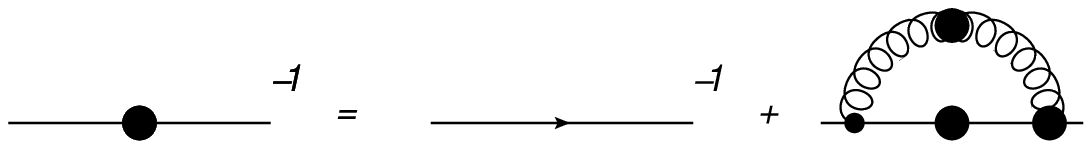}}\hspace{5mm}
\centerline{\includegraphics[width=\columnwidth]{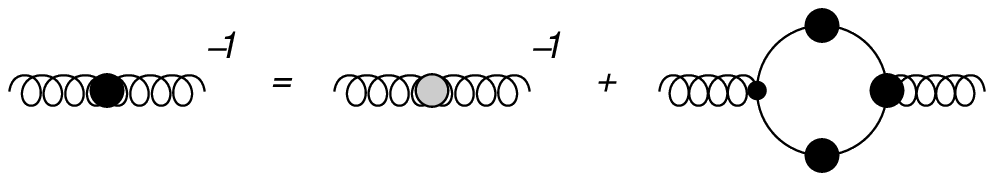}}
\caption{The DSE for the quark and gluon propagators. Filled circles denote 
dressed propagators and vertices. The shaded circle denotes the 
quenched gluon.}
\label{fig:DSE}
\end{figure}

{\bf Dyson-Schwinger equations}\\
The quark propagator is calculated from its Dyson-Schwinger equation (DSE), shown 
in the first line of Fig.~\ref{fig:DSE}.
It depends on the dressed quark-gluon vertex and the dressed gluon propagator, 
which need to be specified. 
For the Landau gauge gluon propagator at finite temperature, the most reliable source up to now
are quenched lattice calculations \cite{Cucchieri:2007ta,Fischer:2010fx}. 
Since we are interested in studying the transition temperatures
and the nature of the transitions at non-vanishing
chemical potential it is mandatory to include quark in-medium effects in
the gluon propagator. To this end we use the lattice data from Ref.~\cite{Fischer:2010fx}
for the quenched gluon and add the quark polarization tensor from the
gluon DSE to effectively unquench the system, see second line of Fig.~\ref{fig:DSE}.
This procedure is not entirely selfconsistent. In particular 
unquenching effects in the Yang-Mills part of the gluon-DSE, i.e. in the ghost-loop
and gluon-loop diagrams hidden in the quenched gluon in Fig.~\ref{fig:DSE} are neglected. 
In the vacuum we compared these effects with the fully selfconsistent treatment of 
Ref.~\cite{Fischer:2003rp}. We found that the approximation used here overestimates
the unquenching effects, with momentum dependent deviations in the gluon propagator  
below the five percent level. Assuming that this remains so at finite temperature 
and chemical potential one would expect to slightly underestimate the transition 
temperatures perhaps on the 5-10 MeV level. We believe that such a treatment is 
sufficiently accurate for the qualitative study envisaged in this work.

Within the polarization tensor we additionally neglect the quark self-energies
in analogy with corresponding works on finite density-QCD in the DSE formalism 
\cite{Nickel:2006vf}.
This approximation is justified by findings comparing lattice and DSE results
for the quark spectral function \cite{Mueller:2010ah} 
showing that the self-energies are suppressed close
to the transition temperature.
The tensor is then given by  
\beq
\Pi_{\mu\nu}(p) = \frac{Z_{1F} N_f}{2} \is_k\,\, \mathrm{Tr}
\left[S_0(q)\gamma_{\mu}S_0(k)\Gamma^0_{\nu}\right],\label{eq:quark-loop}
\eeq
with $q=p+k$, $\isline_k \equiv \sum_n \int \frac{d^3k}{(2\pi)^3}$, 
the vertex renormalization $Z_{1F}$ and the 
quark-gluon vertex for bare propagators $\Gamma^0_{\nu}(p)=g\gamma_\nu \Gamma(p)$,
discussed below. 

Projected to modes longitudinal (L) and transversal (T) 
to the medium, the gluon DSE reads
\beq
\left(Z_{T,L}(p)\right)^{-1} = \left(Z_{T,L}^{qu.}(p)\right)^{-1} + \Pi_{T,L}(p),\label{eq:gluon}
\eeq
with the quenched and unquenched gluon dressing functions $Z_{T,L}^{qu.}$ and $Z_{T,L}$, 
respectively, and the quark loop $\Pi_{T,L}$. 
Equation (\ref{eq:gluon}) may be evaluated in leading order hard thermal loop 
where we obtain thermal momentum dependent masses
\beq
m^2_{T,L}(p)=\frac{N_f\,\pi\,\alpha_{T,L}(p)}{3}\,\left(T^2+\frac{3\mu^2}{\pi^2}\right)
\eeq
with $\alpha_{T,L}(p)=Z_{1F} g^2 Z_{T,L}(p)\Gamma^0(p)/(Z_2^2\,4\pi)$.

For the quark-gluon vertex $\Gamma_\nu$ at finite temperature no information is available
yet, neither from the lattice nor from functional methods. For the purpose of this study 
we therefore rely on a phenomenological {\it ansatz} that has been developed in earlier 
works, Ref.~\cite{Fischer:2009wc,Fischer:2010fx}. To make this work self-contained we
repeat the vertex here,
\begin{widetext}
\beqa \label{vertexfit}
\Gamma_\nu(q,k,p) \!&=&\! \widetilde{Z}_{3}\!\!\left(\delta_{4 \nu} \gamma_4 
\frac{C(k)+C(p)}{2}
\!+\!  \delta_{j \nu} \gamma_j 
\frac{A(k)+A(p)}{2}
\right)\!\!\left( 							
\frac{d_1}{d_2+q^2} \!			
 + \!\frac{q^2}{\Lambda^2+q^2}
\left(\frac{\beta_0 \alpha(\mu)\ln[q^2/\Lambda^2+1]}{4\pi}\right)^{2\delta}\right)\nonumber
\\&&
\eeqa 
\end{widetext}
where $q=(\vec{q},\omega_q)$ denotes the gluon momentum and $p=(\vec{p},\omega_p)$, 
$k=(\vec{k},\omega_k)$ the
quark and antiquark momenta, respectively. Furthermore $2\delta = -18 N_c/(44 N_c - 8 N_f)$ 
is the anomalous 
dimension of the vertex. This 
ansatz contains the quark dressing functions $A$ and $C$ in a manner dictated by 
the Slavnov-Taylor identity of the vertex. It therefore implicitly depends on temperature 
and chemical potential in a meaningful way and also contains unquenching effects via the
corresponding changes in $A$ and $C$. In the quark-DSE we keep this dependence, whereas 
in the gluon polarization we set $A=C=1$ to be consistent with the HTL approximation 
described above.

The parameters $d_1$,$d_2$ and $\Lambda$ of the vertex have been fixed
in Ref.~\cite{Fischer:2010fx} for the case of quenched QCD. The temperature
independent scale $\Lambda$ is associated with the scale of the lattice data for the 
gluon propagator and given by $\Lambda = 1.4$ GeV. The parameter $d_2$ is associated
with the transition from the logarithmic ultraviolet running of the vertex towards the
infrared constant and is given by $d_2 = 0.5 \,\mbox{GeV}^2$. Moreover, $d_1= 4.6 \,\mbox{GeV}^2$ 
parametrizes the infrared strength of the vertex. Both, $d_1$ and $d_2$ have been fixed 
in Ref.~\cite{Fischer:2010fx} such that the chiral and deconfinement transition temperatures
extracted from the quenched quark propagator matched the deconfinement transition temperature
taken from the Yang-Mills lattice input. Note that this procedure is not very sensitive, i.e.
it does not require any fine-tuning: once $d_1$ and $d_2$ are in the right ballpark, 
the resulting transition temperatures have been found to be insensitive to variations
of $d_1$ and $d_2$ up to the ten percent level. Thus once the vertex is chosen such that
it roughly matches the lattice gluon propagator, its details are not so relevant. This is
certainly reassuring. Using a bare quark mass of $m(\mu=30 \,\mbox{GeV})=3.7$ MeV we also
determined the resulting (quenched) pion mass at zero temperature, which is roughly
ten percent above its physical value and independent of changes in $d_1$ and $d_2$. 

In our unquenched calculation we use the same vertex parameters as in the quenched case.
This choice is only justified by simplicity. In general, one also expects unquenching 
effects in the vertex, which may affect the transition temperatures. One way to determine 
the corresponding changes in $d_1$ and $d_2$ would be to calculate unquenched
observable like the pion mass and decay constant in the $T=0$ limit and match to
experiment. Unfortunately this cannot be done within the HTL approximation of the
quark polarization used in this work, since it only contains temperature dependent
fluctuations which vanish in the $T=0$ limit. A more refined treatment of the quark
backreaction including also quantum fluctuations is under way and will be presented in
a future work. We do, however, not expect that unquenching effects in the vertex will 
have a large impact on the transition temperatures. At zero temperature and with fixed
gluon propagator these effects have been found to reduce light hadron masses on the ten 
percent level \cite{Fischer:2008wy}. Thus with fixed gluon propagator one may expect that 
unquenching effects in the vertex will reduce the transition temperatures by a similar 
amount. This is, however, counterbalanced by a corresponding decrease of quark polarization 
effects in the gluon propagator, which in turn increases the transition temperatures again.
Thus the combined effect may indeed be very small.

\begin{figure}[t!]
\centerline{\includegraphics[width=0.9\columnwidth]{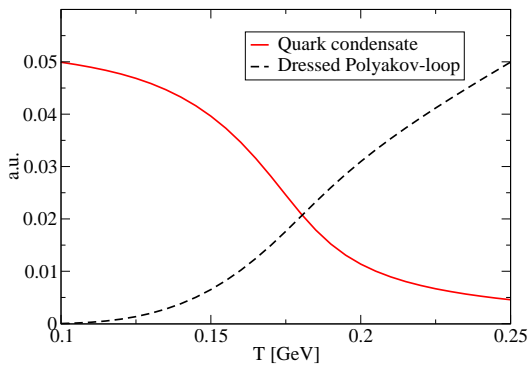}}
\caption{The quark condensate as well as the dressed Polyakov-loop
as a function of temperature in two-flavour QCD.}
\label{fig:finiteT}
\end{figure}

\newpage
{\bf Results}\\
As a first check of the employed truncation scheme we determine the transition temperatures
at $\mu = 0$. In the quenched system, i.e. without the quark-loop contribution of 
Eq.~(\ref{eq:quark-loop}), our system of equations reduces to the 
one outlined in Ref.~\cite{Fischer:2010fx} and we obtain 
$T_c^{N_f=0}= T_{deconf}^{N_f=0} = 277$ MeV. This is in accordance with lattice
caluclations. When we switch from $N_f=0$ to $N_f=2$ the main effect of the 
quark loop onto the gluon propagator is to decrease its strength and to lower the 
characteristic scale $\Lambda_{QCD}$. This also results in decreased chiral and 
deconfinement transition temperatures. We obtain $
T_c^{N_f=2} = 180 \pm 5 \mbox{MeV}$ and $T_{deconf}^{N_f=2} = 195 \pm 5 \mbox{MeV}$ 
which is again close to the corresponding values on the lattice \cite{Karsch:2000kv}.
Here the error bar reflects possible numerical uncertainties in our calculation.
The additional sources for systematic errors of our calculation have been discussed
in the last section. As a rough guidance for the reader we guesstimate an additional 
systematic error of $\pm 20 \mbox{MeV}$. 

Our results for the quark condensate and the dressed Polyakov-loop are shown in 
Fig.~\ref{fig:finiteT}. In the quenched calculation of Ref.~\cite{Fischer:2010fx} a 
sharp rise and fall of the quark condensate close to $T_c$ has been observed, which 
has been attributed to a corresponding behavior of the electric part of the 
gluon propagator. Here, the backreaction of the quark-loop has  
suppressed this effect and the condensate decreases monotonically as expected.
This change of behavior together with the correct order of the transition temperatures gives
us confidence, that our implementation of the backreaction of the
quarks onto the Yang-Mills sector serves its purpose.

\begin{figure}[t!]
\centerline{\includegraphics[width=1.2\columnwidth]{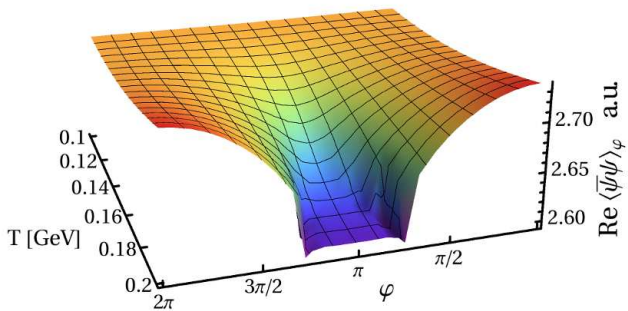}}
\caption{The quark condensate 
$\langle\bar{\psi}\psi\rangle_\varphi$ as a function of temperature and 
boundary angle $\varphi$ at fixed quark chemical potential $\mu=100$ MeV.}
\label{fig:condensateTphi}
\end{figure}

We then switch on the chemical potential and determine the chiral condensate 
as well as the dressed and the conjugate dressed Polyakov loop in the $(T,\mu)$-plane. Note
that for $\mu \ne 0$, $\langle\bar{\psi}\psi\rangle_\varphi$ developes an imaginary part, which 
will be discussed below. A typical example for the behavior of the real part of the 
$\varphi$-dependent quark condensate at a moderate value of $\mu$ is shown in 
Fig.~\ref{fig:condensateTphi}. At $\varphi=\pi$ we observe the chiral cross-over with 
temperature indicated by the ordinary quark condensate. At $\varphi=0,2\pi$ the condensate
rises with temperature similar to $\mu=0$ as discussed in Ref.~\cite{Fischer:2009wc}. 
The deconfinement transition is observed along the 
$\varphi$-direction, when the constant behavior of $\langle\bar{\psi}\psi\rangle_\varphi$
below $T_c$ changes into the typical variations observed above $T_c$ \cite{Fischer:2009wc}.
The new element at finite $\mu$ is the behavior for even larger values of $T$: There 
$\langle\bar{\psi}\psi\rangle_\varphi$ develops an additional discontinuous structure
along the $\varphi$-direction. We have checked that this structure is genuine wrt. variations
of our vertex ansatz. For increased chemical potential the temperature where this discontinuity 
opens moves closer to the deconfinement transition temperature. This suggests that the 
deconfinement transition may turn from crossover to second or first order at chemical potentials 
as large as the one of the chiral critical endpoint (see below). Within the accuracy
of our numerical calculations we could, however, not yet resolve this issue in detail.
This will be addressed in future work. 
The imaginary part of $\langle\bar{\psi}\psi\rangle_\varphi$, not shown
in the figure, is antisymmetric wrt. to $\varphi \rightarrow 2\pi-\varphi$. Therefore
$\Sigma_{\pm 1}=\int_\varphi \left\{Re\langle\bar{\psi}\psi\rangle_\varphi \cos(\varphi)
\pm Im\langle\bar{\psi}\psi\rangle_\varphi \sin(\varphi)\right\}$ stays real and the difference
between quark and anti-quark loops is generated by $Im\langle\bar{\psi}\psi\rangle_\varphi$.

\begin{figure}[t!]
\centerline{\includegraphics[width=\columnwidth]{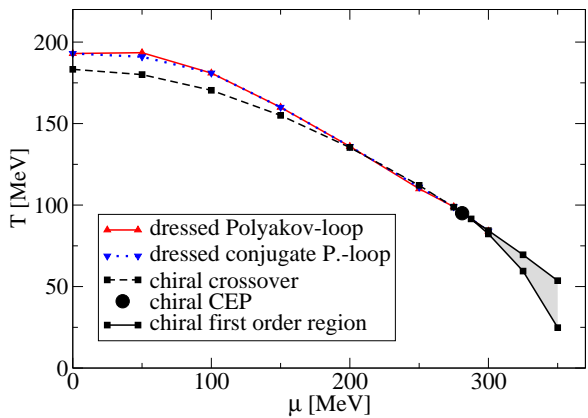}}
\caption{The phase diagram for chiral symmetry breaking ($\chi$) and deconfinement
  of quarks ($\Sigma_1$) and antiquarks ($\Sigma_{-1}$).}
\label{fig:phaseDiag}
\end{figure}
In Fig.~\ref{fig:phaseDiag}, we display the main result of this work: the phase diagram 
of two-flavour QCD. We find a chiral crossover at small 
chemical potential. The deconfinement crossover happens at somewhat larger temperatures
and one observes a characteristic (albeit small) splitting of the 
transition temperatures from the Polyakov-loop and its conjugate \cite{Herbst:2010rf}.
At larger chemical potential all transitions come together again. 
The chiral crossover line goes over into a critical point at 
approximately $(T_{EP},\mu_{EP}) \approx (95,280)$ MeV, followed by the coexistence 
region of a first order transition. Thus we  
find the comparatively large value $\mu_{EP}/T_{EP} \approx 3$. We have checked,
that these values are not overly sensitive to the details of our truncation: when
changing the parameters in the vertex ansatz within a reasonable range we observe
variations of $(T_{EP},\mu_{EP})$ of the order of ten percent. Thus a firm 
conclusion of the present approach seems to be that $\mu_{EP}/T_{EP} \gg 1$: If there
is a critical endpoint, it happens at large chemical potential. This statement
agrees with the result of corresponding calculations in the PQM model, once
quantum corrections have been taken into account \cite{Herbst:2010rf}. 
Expectations from recent lattice calculations at $N_f=2+1$ close to the continuum limit also 
seem to point in this direction \cite{Endrodi:2011gv}.

Certainly, at such large values of the chemical potential, our truncation scheme
may no longer be reliable: baryon effects that are not implicitly included in our 
truncation of the quark-gluon interaction may play an important role here. Also the 
formation of inhomogeneous chiral condensates may be favored upon the homogeneous 
one studied here. We believe that our work provides a suitable basis for further
investigations in these directions.

\vspace*{1mm}
{\bf Acknowledgements}\\
We thank Daniel Mueller and Jan Pawlowski for 
discussions. This work has been supported by the Helmholtz Young 
Investigator Grant VH-NG-332 and the Helmholtz International Center 
for FAIR within the LOEWE program of the State of Hesse.

\end{document}